\documentclass[aps,prc,superscriptaddress,twoside,twocolumn,nofootinbib,%
showpacs,floatfix]{revtex4-1}

\usepackage{amsmath,amssymb}
\usepackage{graphicx,bm}
\usepackage{slashed}

\usepackage{ulem} 
\usepackage[usenames]{color}

\renewcommand\sout{\bgroup \color{red} \ULdepth=-.5ex \ULset}

\begin{document}

\title{QCD sum rule study on the $f_0(980)$ structure as a pure $K \bar{K}$ bound state}

\author{Hee-Jung Lee}
\email{hjl@cbnu.ac.kr}
\affiliation{Department of Physics Education, Chungbuk National University,
Cheongju, Chungbuk 361-763, Korea}

\author{N. I. Kochelev}
\email{kochelev@theor.jinr.ru}
\affiliation{Department of Physics, Kyungpook National University,
Daegu 702-701, Korea}
\affiliation{Bogoliubov Laboratory of Theoretical Physics,
Joint Institute for Nuclear Research, Dubna, Moscow region, 141980 Russia}

\author{Yongseok Oh}
\email{yohphy@knu.ac.kr}
\affiliation{Department of Physics, Kyungpook National University,
Daegu 702-701, Korea}
\affiliation{Asia Pacific Center for Theoretical Physics, Pohang, Gyeongbuk 790-784, Korea}

\date{\today}
\begin{abstract}
We perform a QCD sum rule analysis for the scalar $f_0(980)$ meson to investigate whether
it can be described as a pure bound state of $K$ and $\bar{K}$ mesons.
Based on the QCD sum rule with the operators of up to dimension 10 within the operator
product expansion, we found that it is hard to treat the $f_0(980)$ as a simple $K\bar{K}$
bound state, which implies that the $f_0(980)$ scalar meson has more complicated structure
being mixed states of various configurations.
\end{abstract}


\maketitle



The structure of the scalar meson nonet has been a long-standing puzzle in hadron physics.
It is now widely accepted that the simplest picture, where the scalar mesons are described
as orbital excitations of quark-antiquark pairs, is not compatible with
the experimental observations on the decay modes and mass spectra~\cite{AT04}.
This led to the idea that these scalar mesons are crypto-exotic tetraquark
states~\cite{Jaffe77a},
and there have been a lot of studies along this direction.
Depending on the details of the structure of the tetraquark states, the scalar mesons
are considered as diquark-antidiquark bound
states~\cite{Jaffe04,MPPR04a,BNNB05,WY05,CHZ0607,MACD06},
two-meson molecular states~\cite{WI8290,CIK93,AGS97,BHHKK04,BGL08,SBL12},
or hybrid states~\cite{AADM05}. (See also Ref.~\cite{BHS07}.)

Among the low-lying scalar mesons, the $f_0(980)$ attracts much interests since
the seminal work of Weinstein and Isgur, which investigated the $f_0(980)$ as a $K\bar{K}$
molecular state~\cite{WI8290}.
In a recent work~\cite{BGL08}, for example, the properties of the $f_0(980)$ were reanalyzed
in a phenomenological Lagrangian approach assuming a pure $K{\bar K}$ bound state,
and the calculated decay widths for $f_0(980) \to \pi\pi$ and
$f_0(980) \to \gamma\gamma$ were claimed to be consistent with the available data.
In a recent work~\cite{SBL12}, however, the scalar and isoscalar meson resonances are
investigated in various channels of $\pi\pi$ scattering, which raised the possibility of
the $f_0(980)$ as a pure $\eta\eta$ bound state rejecting the pure $K\bar{K}$ structure.
All these ambiguities show that the structure of scalar mesons is nontrivial and
more QCD-based approaches are required for understanding the structure of scalar mesons.

The QCD sum rule (QCDSR) approach is known to be one of the ways to investigate the
hadron properties from QCD in a direct way~\cite{SVZ79}.
This approach was used to study the diquark picture of scalar mesons~\cite{WY05}, and
it was recently shown by one of us that the QCDSR does not support the picture of the
$f_0(980)$ as a pure $\eta\eta$ bound state~\cite{lee11}.
In the present work, we construct the QCDSR for the $f_0(980)$ to test whether it can be
described as a pure $K{\bar K}$ bound state.
To this end, we obtain the QCDSR up to dimension $d=10$ operators within the operator
product expansion (OPE).

The wave function of the $f_0(980)$ meson as a pure $K\bar{K}$ bound state
is written generally as
\begin{equation}
|f_0 (980)\rangle=\alpha |K^+ K^-\rangle +\beta |K^0 \bar{K}^0\rangle.
\end{equation}
With the following $K$ meson interpolating currents,
\begin{eqnarray}
&&
J_{K^+}=i\bar{s}\gamma_5 u,\quad
J_{K^-}=i\bar{u}\gamma_5 s,
\nonumber \\ &&
J_{K^0}=i\bar{s}\gamma_5 d,\quad
J_{\bar{K}^0}=i\bar{d}\gamma_5 s ,
\end{eqnarray}
the interpolating current for the $f_0(980)$ in QCDSR approach becomes
\begin{eqnarray}
J_{f_0}&=&\alpha J_{K^+}J_{K^-} +\beta J_{K^0}J_{\bar{K}^0}
\nonumber\\
&=&
-\left[ \alpha(\bar{s}\gamma_5u)(\bar{u}\gamma_5s) +
\beta(\bar{s}\gamma_5d)(\bar{d}\gamma_5s) \right].
\label{current}
\end{eqnarray}
Then the vacuum expectation value of the time ordered product of the currents reads
\begin{widetext}
\begin{eqnarray}
\langle 0| TJ_{f_0}(x)J^\dagger_{f_0} (0)|0\rangle&=&
\langle 0|
T\bigg\{ \alpha^2[\bar{s}(x)\gamma_5u(x)][\bar{u}(x)\gamma_5s(x)]
[\bar{s}(0)\gamma_5u(0)][\bar{u}(0)\gamma_5s(0)]
\nonumber\\ && \hspace{-2cm} \mbox{}
+\alpha\beta [\bar{s}(x)\gamma_5u(x)][\bar{u}(x)\gamma_5s(x)]
[\bar{s}(0)\gamma_5d(0)][\bar{d}(0)\gamma_5s(0)]
+\alpha\beta [\bar{s}(x)\gamma_5d(x)][\bar{d}(x)\gamma_5s(x)]
[\bar{s}(0)\gamma_5u(0)][\bar{u}(0)\gamma_5s(0)]
\nonumber\\ && \hspace{-2cm} \mbox{}
+\beta^2 [\bar{s}(x)\gamma_5d(x)][\bar{d}(x)\gamma_5s(x)]
[\bar{s}(0)\gamma_5d(0)][\bar{d}(0)\gamma_5s(0)]
\bigg\} |0\rangle .
\label{top}
\end{eqnarray}
Since the disconnected terms do not contribute to the QCSSR, here we present only
the connected terms.
Then the first term can be transformed as
\begin{eqnarray}
\!\!\!\!\!
\langle 0| T [\bar{s}(x)\gamma_5u(x)][\bar{u}(x)\gamma_5s(x)]
[\bar{s}(0)\gamma_5u(0)][\bar{u}(0)\gamma_5s(0)]|0\rangle
=
{\rm Tr}[S_s^{ba'}(x,0)\gamma_5S_u^{a'b}(0,x)\gamma_5]
{\rm Tr}[S_s^{b'a}(0,x)\gamma_5S_u^{ab'}(x,0)\gamma_5]  ,
\label{1st}
\end{eqnarray}
\end{widetext}
in terms of the quark propagator $S_{q}^{ab}(x,y)$ with the color indexes $a,b$.
One can easily verify that, in Eq.~(\ref{top}), replacing the quark flavor $u$ in the first
two terms by $d$ yields the last two terms.
Since we are working in the chiral limit $m_u=m_d=0$, the first two terms and
the last two terms give the same contribution.
Furthermore, the second and the third terms have disconnected diagrams only,
which leads to the overall factor $\alpha^2+\beta^2$ in Eq.~(\ref{top}).

\begin{figure*}[t]
\centering
\includegraphics[width=350 pt]{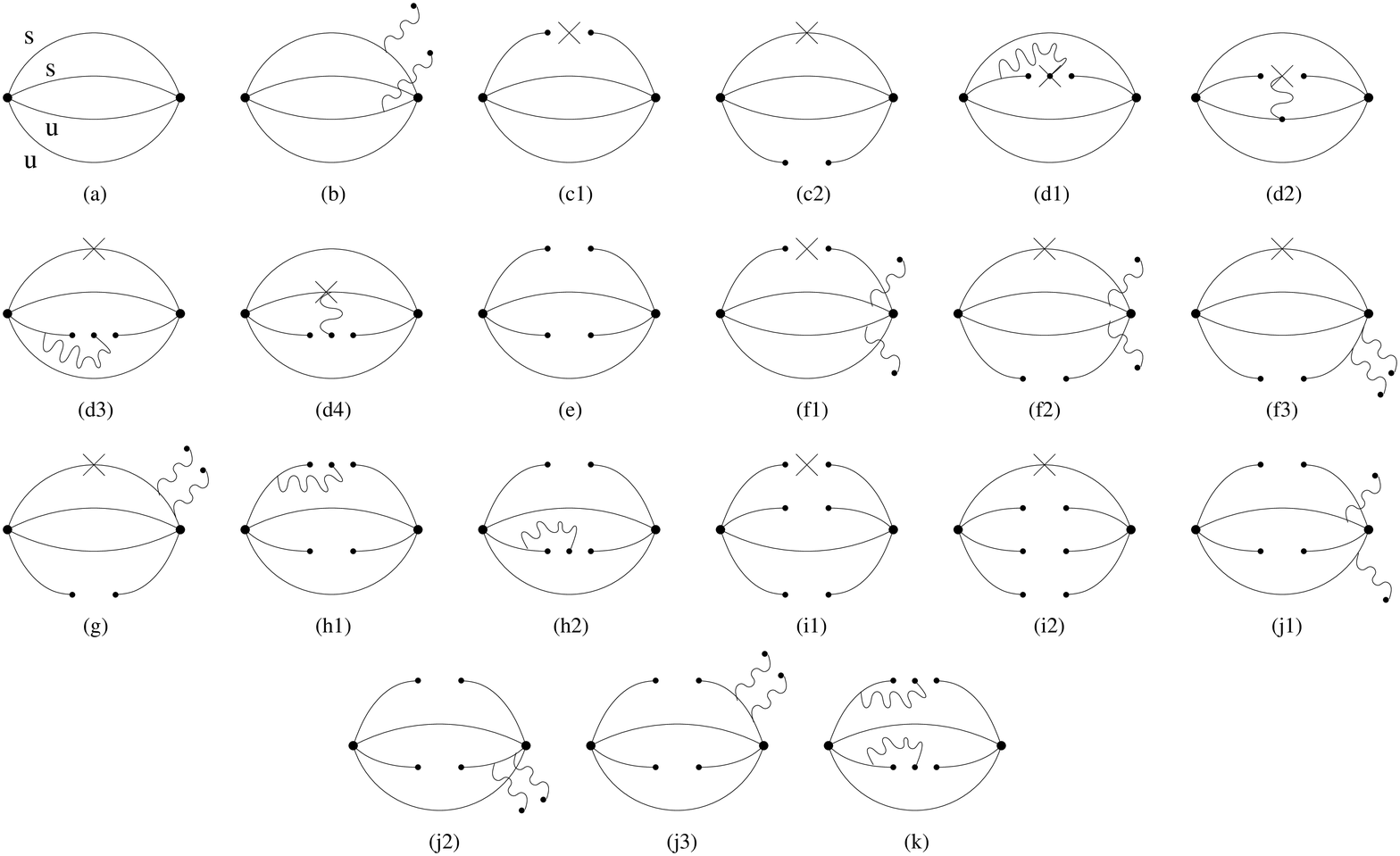}
\caption{Diagrammatic representations of the terms in Eq.~(\ref{Im}).
Upper two lines correspond to the $s$ quark and lower two lines
to the $u$ quark. The symbol $\times$ denotes the strange quark mass $m_s$.
Here only the nonvanishing diagrams are shown.}
\label{fig:Diag}
\end{figure*}

The correlator
$\Pi(q^2)=i\int d^4 x\, e^{iq\cdot x} \langle 0 | TJ_{f_0}(x)J^\dagger_{f_0}(0)
|0 \rangle $ can then be calculated within the OPE up to $O(m_s)$ and $O(g_c^2)$
keeping the operators of dimension up to 10.
By making use of the quark propagator of Ref.~\cite{LKV06}, the imaginary part of
the correlator is obtained as
\begin{widetext}
\begin{eqnarray}
\frac{1}{\pi}\, \mbox{Im}\, \Pi^{\rm OPE}(q^2) &=& \left( \alpha^2 + \beta^2 \right)
\bigg[ \frac{1}{2^{14} \left( 5\pi^6 \right) } (q^2)^4
+\frac{g_c^2\langle G^2\rangle}{2^{12} \pi^6 }(q^2)^2
+\frac{m_s}{2^8\pi^4}\left( \langle \bar{s}s\rangle-2\langle \bar{u}u\rangle
\right)(q^2)^2
\nonumber\\ && \mbox{}
+\frac{m_s}{2^8\pi^4} \left\{ 2ig_c\langle \bar{s}\sigma\cdot G s\rangle
+3ig_c\langle \bar{u}\sigma\cdot G u\rangle \right\} q^2
\nonumber\\ && \mbox{}
+\frac{m_s i g_c \langle{\bar u}\sigma\cdot G u\rangle}{2^7\pi^4}
q^2 \left\{ -2\ln(q^2/\Lambda^2)+\ln\pi+\psi(3)+\psi(2)+2\gamma_{\rm EM}^{} \right\}
+\frac{\langle\bar{u}u\rangle\langle\bar{s}s\rangle}{2^4\pi^2}q^2
\nonumber\\ && \mbox{}
+\frac{m_sg_c^2\langle G^2\rangle}{2^9\pi^4} \left\{ \langle \bar{s}s\rangle
-\frac{2^3}{3}\langle \bar{u}u\rangle \right\}
-\frac{m_sg_c^2\langle G^2\rangle \langle \bar{u} u\rangle}{2^8\pi^4}
\left\{ -2\ln(q^2/\Lambda^2)+\ln\pi+\psi(2)+\psi(1)+2\gamma_{\rm EM}^{} -\frac23 \right\}
\nonumber\\ && \mbox{}
-\frac{1}{2^5\pi^2} \left(\langle\bar{u}u\rangle ig_c\langle\bar{s}\sigma\cdot G s\rangle
+\langle\bar{s}s\rangle ig_c\langle\bar{u}\sigma\cdot G u\rangle \right)
-\frac{m_s\langle \bar{u}u\rangle\langle \bar{s}s\rangle}{2\cdot3}
\left(\langle \bar{u}u\rangle-\frac{1}{2}\langle \bar{s}s\rangle \right)\delta(q^2)
\nonumber\\ && \mbox{}
+\frac{5g_c^2\langle G^2\rangle \langle \bar{u}u\rangle\langle \bar{s}s\rangle}{2^6\cdot3^2\pi^2}
\, \delta(q^2)
+\frac{13ig_c\langle \bar{u}\sigma\cdot G u\rangle ig_c\langle \bar{s}\sigma\cdot G s\rangle}
{2^{9}\cdot3\pi^2}\, \delta(q^2)\bigg] ,
\label{Im}
\end{eqnarray}
\end{widetext}
where $g_c$ is the strong coupling constant and
$\psi(n) = 1 + 1/2 + \dots + 1/(n-1) - \gamma_{\rm EM}^{}$
with the Euler-Mascheroni constant $\gamma_{\rm EM}^{}$.
Here, we have used the factorization hypothesis in calculating the condensates of the operators
of dimension higher than 6.
The diagrammatic representation of each term is shown in Fig.~\ref{fig:Diag}.

Decomposing the spectral sum, which is generated from the dispersion relation of
the correlator, into a narrow single resonance and the continuum, and applying the
hadron-quark duality hypothesis with the Borel transform as well, we have the
following sum rule,
\begin{equation}
\frac{1}{\pi}\int_0^{s_0^2}ds^2 e^{-s^2/M^2}\, \mbox{Im}\, \Pi^{\rm OPE}(s^2)
=2f_{f_0}^2m_{f_0}^8e^{-m_{f_0}^2/M^2} ,
\end{equation}
with the convention
$\langle 0 | J_{f_0}(0) | f_0 (980) \rangle = \sqrt{2} f_{f_0} m_{f_0}^4$.
Here, $s_0^{}$ and $M$ denote the threshold for the continuum and the Borel mass, respectively.
The imaginary part of the correlator in Eq.~(\ref{Im}) gives the explicit QCDSR for the $f_0(980)$ as
\begin{widetext}
\begin{eqnarray}
2f_{f_0}^2m_{f_0}^8e^{-m_{f_0}^2/M^2} &=&
(\alpha^2+\beta^2)\bigg[\frac{3}{2^{11} \left( 5\pi^6 \right) }M^{10}E_4 (M^2)
+\frac{g_c^2\langle G^2\rangle}{2^{11}\pi^6}M^6E_2(M^2)
+\frac{m_s}{2^7\pi^4}\bigg\{ \langle \bar{s}s\rangle-2\langle \bar{u}u\rangle\bigg\} M^6E_2(M^2)
\nonumber\\&& \mbox{}
+\frac{m_s}{2^8\pi^4}\bigg\{ 2ig_c\langle \bar{s}\sigma\cdot G s\rangle
+3ig_c\langle \bar{u}\sigma\cdot G u\rangle\bigg\} M^4E_1(M^2)
+\frac{m_sig_c\langle{\bar u}\sigma\cdot G u\rangle}{2^7\pi^4}M^4\overline{W}_1(M^2)
\nonumber\\&& \mbox{}
+\frac{\langle\bar{u}u\rangle\langle\bar{s}s\rangle}{2^4\pi^2}M^4E_1(M^2)
+\frac{m_sg_c^2\langle G^2\rangle}{2^9\pi^4}\bigg\{ \langle \bar{s}s\rangle
-\frac{2^3}{3}\langle \bar{u}u\rangle\bigg\} M^2E_0(M^2)
-\frac{m_sg_c^2\langle G^2\rangle \langle \bar{u} u\rangle}{2^8\pi^4}M^2W_0(M^2)
\nonumber\\ && \mbox{}
-\frac{1}{2^5\pi^2}\bigg\{ \langle\bar{u}u\rangle ig_c\langle\bar{s}\sigma\cdot G s\rangle
+\langle\bar{s}s\rangle ig_c\langle\bar{u}\sigma\cdot G u\rangle\bigg\} M^2E_0(M^2)
-\frac{m_s\langle \bar{u}u\rangle\langle \bar{s}s\rangle}{2^2\cdot3}
\bigg\{ \langle \bar{u}u\rangle-\frac{1}{2}\langle \bar{s}s\rangle\bigg\}
\nonumber\\ && \mbox{}
+\frac{5g_c^2\langle G^2\rangle \langle \bar{u}u\rangle\langle \bar{s}s\rangle}{2^7\cdot3^2\pi^2}
+\frac{13ig_c\langle \bar{u}\sigma\cdot G u\rangle ig_c\langle \bar{s}\sigma\cdot G s\rangle}
{2^{10}\cdot3\pi^2}\bigg],
\label{SR}
\end{eqnarray}
where we have used $\langle {\bar u}u\rangle=\langle {\bar d}d\rangle$
in the chiral limit and
\begin{eqnarray}
E_n (M^2) &=& \frac{1}{\Gamma(n+1)M^{2n+2}} \int_0^{s_0^2} ds^2\, e^{-s^2/M^2}
\left( s^2 \right)^n ,
\nonumber \\
\overline{W}_n (M^2) &=& \frac{1}{\Gamma(n+1)M^{2n+2}} \int_0^{s_0^2} ds^2\, e^{-s^2/M^2}
\left( s^2 \right)^n
\left\{ -2\ln(s^2/\Lambda^2)+\ln\pi+\psi(n+1)+\psi(n+2)+2\, \gamma_{\rm EM}^{} \right\},
\end{eqnarray}
with $W_n(M^2) = \overline{W}_n (M^2) - \frac23 E_n (M^2)$.
\end{widetext}

\begin{figure*}[t]
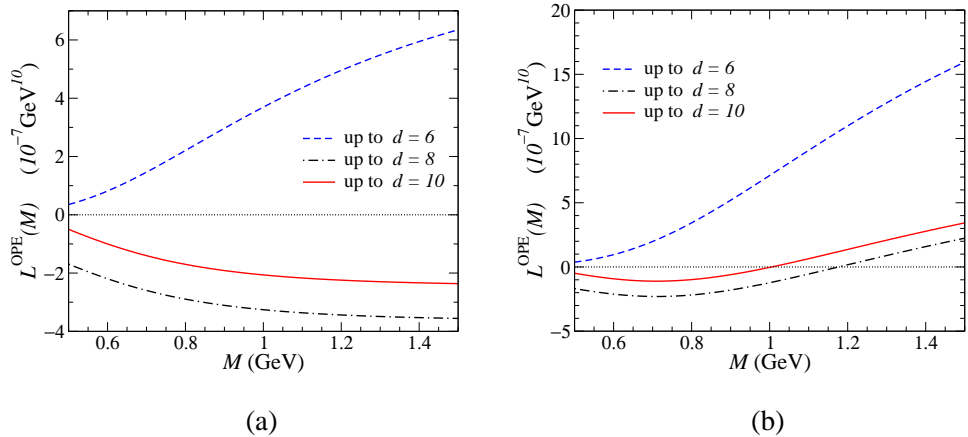

\centering
\includegraphics[width=170 pt]{fig2a.eps}
\qquad \includegraphics[width=170 pt]{fig2b.eps}
\caption{$L^{\rm OPE}$ as a function of $M$ for
(a) $s_0^{}=1.2$~GeV and (b)
$s_0^{} = 1.5$~GeV.
Dashed and dot-dashed lines are $L^{\rm OPE}$ obtained with
the operators of up to dimension 6 and 8, respectively.
The solid lines show the full calculation of up to dimension~10.}
\label{fig:LOPE}
\end{figure*}

For numerical analysis, we use the standard values and relations for $m_s$ and the
condensates as
\begin{eqnarray}
&&
\langle\bar{u}u\rangle=-(0.25)^3 \mbox{ GeV}^3,\quad
\langle\bar{s}s\rangle=f_s\langle\bar{u}u\rangle,
\nonumber\\ &&
\langle g_c^2G^2\rangle =
0.5 \mbox{ GeV}^4, \quad
m_s = 0.15 \mbox{ GeV},
\nonumber\\ &&
ig_c\langle\bar{u}\sigma\cdot Gu\rangle =
0.8 \mbox{ GeV}^2 \langle \bar{u}u \rangle,
\nonumber \\ &&
ig_c\langle\bar{s}\sigma\cdot Gs\rangle =
f_s ig_c\langle\bar{u}\sigma\cdot Gu \rangle
\end{eqnarray}
with $f_s=0.8$ and $\Lambda = 0.5$~GeV.
Since the QCDSR is proportional to $\alpha^2+\beta^2$, the results are independent of the
choice on $\alpha$ and $\beta$.

\begin{figure}[t]
\includegraphics[width=170 pt]{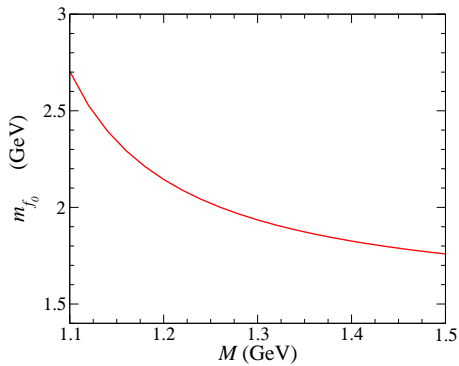}
\caption{The fitted mass from the sum rule (\ref{SR}) as a function of $M$
for $s_0^{}=1.5$~GeV.}
\label{fig:mass}
\end{figure}

Defining the right hand side of the sum rule in Eq.~(\ref{SR}) by $L^{\rm OPE}(M)$, we analyze
its behavior as a function of the Borel mass $M$.
Shown in Fig.~\ref{fig:LOPE} is $L^{\rm OPE}(M)$ for the threshold $s_0^{}=1.2$~GeV and
$1.5$~GeV.
Here, the dashed, dot-dashed, and solid lines correspond to $L^{\rm OPE}(M)$ with
the operators of $d \le 6$, $d \le 8$, and $d \le 10$, respectively.
This shows that the contribution from the operators of dimension 8 to the QCDSR
is large and negative for both cases.
For $s_0^{}=1.2$~GeV, in contradiction with a positive definite value of the left hand side
of Eq.~(\ref{SR}), the large negative contribution from the operators of dimension 8
makes the full $L^{\rm OPE}(M)$ have a definite negative value in the physical Borel region
less than the threshold.
This is similar to the result found in Ref.~\cite{Lee06}, where the QCDSR for the light scalar meson
nonet was analyzed by assuming the scalar diquark-antidiquark structure.
For $s_0^{}=1.5$~GeV, the contributions from the operators of dimension 6 and 10 are large enough
to overcome the negative contribution from the dimension 8 operators in
the Borel region $M \ge 1$~GeV.
However, as shown in Fig.~\ref{fig:mass}, it is difficult to find the Borel window,
where the fitted mass does not have strong dependence on $M$.
Furthermore, the fact that the fitted mass is larger than the value of the threshold is
in contradiction with the basic concept of the QCDSR.
In addition, the ratio of the pole to continuum contributions is found to be very small ($\sim 0.03$),
which violates one of the main requirements to have a reliable QCDSR as discussed in Ref.~\cite{Bracco12}.

We have also tested the sum rule with $s_0^{} > 1.5$~GeV to find that the Borel region of positive
$L^{\rm OPE}(M)$ becomes wider.
However, the fitted mass is very high (about 1.8~GeV for $s_0^{}=2.0$~GeV, for example)
compared to the $f_0(980)$ mass.
These observations lead us to conclude that it is hard to consider the $f_0(980)$ as a pure
$K{\bar K}$ bound state.
We also point out that the possible strong deviations of the values of the condensates
of dimension 6 and 8 from the factorization hypothesis in the level presented in 
Ref.~\cite{Narison05} does not change our main conclusion.

In summary, we have constructed and analyzed the QCDSR within the OPE with the operators of
$d \le 10$ by assuming the pure $K{\bar K}$ structure for the $f_0(980)$.
Our analyses show that there is no value of the threshold which guarantees the positivity
of $L^{\rm OPE}$ and weak dependency of the fitted mass for the $f_0(980)$
on the Borel mass simultaneously.
This leads to the conclusion that the $f_0(980)$ has a very complicated structure other than
a pure $K{\bar K}$ state.
Therefore, it would be interesting to investigate an admixture of four quark configurations
and two quark configuration for the internal structure of the $f_0(980)$.

\bigskip

We are grateful to S.~B. Gerasimov for fruitful discussions on various aspects
of this research.
Y.O. thanks H.~Kim for valuable comments.
This work was supported in part by the National Research Foundation funded
by the Korean Government (Grant No.\ NRF-2011-220-C00011 and
Grant No.\ NRF-2010-0009381).
The work of N.K. was supported in part by the MEST of the Korean Government
(Brain Pool Program No.\ 121S-1-3-0318).
We also acknowledge that this work was initiated through the series of
APCTP-BLTP JINR Joint Workshop.

\end{document}